\documentclass[aps,prl,twocolumn,groupedaddress,showpacs,showkeys,amsmath,amssymb]{revtex4}

\usepackage{amsfonts}
\usepackage{amssymb,amsmath}
\usepackage{graphicx}
\usepackage{dcolumn}
\usepackage{bm}

\newcommand{\be}{\begin{equation}}
\newcommand{\ee}{\end{equation}}
\newcommand{\bea}{\begin{eqnarray}}
\newcommand{\eea}{\end{eqnarray}}
\newcommand{\la}{\langle}
\newcommand{\ra}{\rangle}
\newcommand{\ld}{\left(}
\newcommand{\rd}{\right)}
\newcommand{\lb}{\left\{}
\newcommand{\rb}{\right\}}
\newcommand{\lbr}{\left[}
\newcommand{\rbr}{\right]}
\newcommand{\non}{\nonumber \\ }

\begin{document}

\title{Exact results for spatial decay of the one-body density matrix\\
in low-dimensional insulators }

\author{Janusz J\c{e}drzejewski$^{1,2}$}
\email[Electronic address:]{jjed@ift.uni.wroc.pl}

\author{ Taras Krokhmalskii$^{3,1}$}
\email[Electronic address:]{krokhm@icmp.lviv.ua}

\affiliation{$^{1}$Institute of Theoretical Physics,
        University of Wroc\l aw, 
        pl. Maksa Borna 9, 50-204 Wroc\l aw, Poland\\
        $^{2}$Department of Theoretical Physics,
                    University of \L \'{o}d\'{z}, 
        ul. Pomorska 149/153, 90-236 \L \'{o}d\'{z}, Poland\\
        $^{3}$Institute for Condensed Matter Physics, 
        1 Svientsitskii Str., L'viv-11, 79011, Ukraine}

\date{\today}

\pacs{ 71.10.Fd, 71.20.-b}
\keywords{tight-binding electrons, band theory, insulators,
one-body density matrix, spatial decay, correlation length}

\begin{abstract}

We provide a tight-binding model of insulator, for which we derive
an exact analytic form of the one-body density matrix and its
large-distance asymptotics in dimensions $D=1,2$. The system is
built out of a band of single-particle orbitals in a periodic
potential. Breaking of the translational symmetry of the system
results in two bands, separated by a direct gap whose width is
proportional to the unique energy parameter of the model. The form
of the decay is a power law times an exponential. We determine the
power in the power law and the correlation length in the
exponential, versus the lattice direction, the direct-gap width,
and the lattice dimension. In particular, the obtained exact
formulae imply that in the diagonal direction of the square
lattice the inverse correlation length vanishes linearly with the
vanishing gap, while in non-diagonal directions, the linear
scaling is replaced by the square root one. Independently of
direction, for sufficiently large gaps the inverse correlation
length grows logarithmically with the gap width.
\end{abstract}

\maketitle

The rapid progress in computational techniques used for
calculating properties of solids enabled researchers to implement a
localized real-space approach to describing such properties. It
has made possible large-scale calculations based on the density
functional theory \cite{payne}, in particular calculating  the
electronic structure of solids by means of $O(N)$ methods
\cite{goed,kohn1}. In all these methods it is the one-body density
matrix (DM) that is the central quantity. Its decay rate
determines the degree of locality of all quantities relevant for
physics and is decisive for the speed of the involved algorithms.
This is why one observes a growing interest in localization
properties of DM in recent years.

However, the first result concerned with the rate of decay of DM
was published by W. Kohn as early as in 1959 \cite{kohn2}. Kohn
demonstrated  that in a one-dimensional centrosymmetric crystal,
the Wannier functions, and hence the DM, decay exponentially with
sufficiently large distance. This result was extended to crystals
of higher dimensionality by J. Des Cloizeaux \cite{cloizeaux}.

For the reasons explained briefly above, the problem of the rate
of decay of DM became the one of vital importance about forty
years later.The typical questions in this respect are concerned
with the dependence of the decay rate on dimensionality and energy
parameters (especially the direct-gap width) of systems under
study. First, Baer and Gordon \cite{baer} gave general arguments
that the exponential decay, discovered by Kohn, is valid
irrespectively of dimensionality, and moreover, the correlation
length, $\xi$, that characterizes the exponential decay, scales
with the direct-gap width, $\delta$, like ${\xi}^{-1} \sim
\sqrt{\delta}$. Then, the question of the rate of decay of DM was
reconsidered by Ismail-Beigi and Arias \cite{beigi} in a general,
model-independent context, for crystals of arbitrary
dimensionality, that is in systems described by single-particle
orbitals in periodic potentials. They found also the exponential
decay law in insulators of arbitrary dimensionality but questioned
the validity of the scaling ${\xi}^{-1} \sim \sqrt{\delta}$,
obtained in \cite{baer}. They argued that instead of the square
root scaling, the linear one, ${\xi}^{-1} \sim \delta$, holds at
least in the limit of vanishing direct-gap width. After that, He
and Vanderbilt \cite{he} discovered a power-law factor, which
multiplies the exponential factor in the asymptotic formula for DM
in one-dimensional crystals. All these findings created a
situation in which it was highly desirable to construct a $D>1$
model of insulator whose DM could be investigated exactly in the
large-distance asymptotic regime, and the dependence of power and
exponential laws of decay on dimensionality and direct-gap width
could be determined unumbiguously.

Quite recently, an attempt at constructing such a model was made
by Taraskin et al \cite{taraskin1}.
They proposed a simple tight-binding model,
built out of two kinds of single-particle orbitals at each lattice
site, which they consider to contain the basic features of an
insulator. They succeeded in demonstrating analytically the
existence of a power-law factor and an exponential one in all
three dimensions $D= 1,2,3$. However, the result was obtained only
for very large values of one of the energy parameters of the
model, which is in no definite relation with the direct-gap width
\cite{jk} (unless some additional assumptions about the energy
parameters of the model are made). Moreover, the model proposed by
Taraskin et al is not a kind of crystal studied in the papers
cited above, since it is translation invariant.

We have succeeded in deriving analytically the exact form of DM
(in terms of special functions) and its large-distance properties
for another system, which is a crystal. The system consists of one
kind of single-particle orbitals at each lattice site, which are
subjected to an external periodic potential. There is the unique
energy parameter that determines the strength of the periodic
potential. This system exhibits two bands separated by a direct
gap whose width is proportional to the unique energy parameter. We
have derived the power-law and exponential factors in the large
distance asymptotics of DM in dimensions $D=1,2$, for small and
large direct-gap widths, and in arbitrary direction in the
$D=2$-case.

Consider the model described by the following second-quantized
Hamiltonian
\be
\label{hcb}
H = \sum_{\bm i}U_{\bm i}a^+_{\bm i}a_{\bm i}
+ t \sum_{\la \bm i,\,\bm j \ra} \ld a^+_{\bm i}a_{\bm j}+h.c.
\rd .
\ee
In the above expression, $\bm i,\bm j$ represent the
lattice sites of a $D$-dimensional lattice, while $\la \bm
i,\bm j \ra$ stands for a pair of nearest neighbors on this
lattice. The operators $\{a^{+}_{\bm i},a_{\bm i} \}$ create,
annihilate, respectively, a spinless fermion in a single-particle
orbital $|\bm i \ra$, belonging to an orthonormal basis.
The value of the external potential at site $\bm j$ is $U_{\bm j}$.
Suppose that the underlying lattice is a $D$- dimensional simple cubic
lattice, which consists of two interpenetrating sublattices
(the even and odd sublattices), such that the nearest neighbors of a site
on one sublattice belong to the other one.
Then, we set $U_{\bm j}= U_1$ on the odd sublattice and $U_{\bm j}=U_2$
on the even one. Under the periodic boundary conditions the Hamiltonian
(\ref{hcb}) is block-diagonalized by the plane wave orbitals $|\bm k \ra$
with the wave vector in the first Brillouin zone of the lattice.
Specifically, shifting the energy scale to $(U_1 + U_2)/2$ and expressing
all the energies in the units of the transfer integral $t$, we obtain
the upper, $\Lambda^+_{\bm k}$,  and the lower, $\Lambda^-_{\bm k}$,
bands of eigenenergies:
\be
\label{cbdisp}
\Lambda^{\pm}_{\bm k}=\pm  2\sqrt{(u/2)^2+S_{\bm k}^2} \equiv \pm
2\Delta(\bm k ),
\ee
In (\ref{cbdisp}) the wave vector ${\bm k}$  is restricted to the
first Brillouin zone of one of the sublattices,
$u\equiv(U_2-U_1)/2t$ is the unique energy parameter of the model,
and
$S_{\bm k}=\frac{1}{2}
\sum_{\bm j;\, \la \bm i,\,\bm j \ra} \exp\ld i{\bm k(\bm i\!-\! \bm j)}\rd$
stands for the structure factor.
Without any loss of generality, we can set $u > 0$. The two
bands are mutually symmetric about zero and are separated by
the gap $\delta=2u$. For $u \neq 0$
and completely filled lower band, the system given by (\ref{hcb})
is an insulator in the sense of the band theory, which we call the
{\it chessboard insulator}.
The eigenvectors corresponding to the eigenvalues $\Lambda^{\pm}_{\bm k}$,
$|\bm k, \pm \ra$, respectively, are linear combinations of the vectors
$|\bm k\ra$ and $|\bm k + \bm \pi \ra$, where $|\bm \pi \ra$ stands for
the vector whose all components are equal to $\pi$.
By means of these eigenvectors one can calculate
the zero-temperature, non-diagonal elements of DM:
\bea
\label{DModd}
\la a^+_{\bm i}a_{\bm i+ \bm r} \ra  \!\! &=&\!\!
  - \frac{1}{2}\sum_{l=1}^{D} {\cal{S}}_l {\cal{R}}(\bm r),
\ \ \mbox {if} \   \sigma_{\bm r}=2m+1, \non
\la a^+_{\bm i}a_{\bm i + \bm r} \ra \!\!&=&\!\!
-\frac{u}{2} (-1)^{\sigma_{\bm i}} {\cal{R}}(\bm r),
 \ \mbox {if} \  \sigma_{\bm r}=2m ,
\eea
where
\be
\label{Rcb}
{\cal{R}}(\bm r)=
(2\pi)^{-D}
\int_{B.Z.}{\mbox d}\bm k
\exp\ld i\bm k \bm r\rd  \Delta^{-1}(\bm k),
\ee
and ${\cal{S}}_1 {\cal{R}}(\bm r)
\equiv {\cal{R}}(r_1 + 1,\ldots,r_D) +
{\cal{R}}(r_1 - 1,\ldots,r_D)$, and so on for $l=2, \ldots, D $.
In (\ref{Rcb}), the D-dimensional integral is taken over the first Brillouin
zone of a sublattice and $\bm r \neq 0$. For the discussion that follows it is
convenient to introduce the parameter $\sigma_{\bm r}\equiv\sum_{l=1}^D |r_l|$,
which amounts to a (noneuclidean) distance between the lattice point ${\bm r}$
and the origin.
Note that the matrix elements of DM depend on ${\cal{R}}(\bm r)$
evaluated only at the points $\bm r$ with $\sigma_{\bm r}$ even.

In the cases studied here we have found
that up to a coefficient independent of $\sigma_{\bm r}$,
\be
\label{asymptot}
{\cal {R}}(\bm r) \sim \sigma_{\bm r}^{-\gamma}
\exp(-\sigma_{\bm r} / \xi ),
\ee
for sufficiently large $\sigma_{\bm r}$, with the power $\gamma$ depending
only on the dimensionality of our insulator and $\xi$ depending on the direction
of $\bm r$ and the direct-gap width.

\begin{figure}
\begin{center}
\includegraphics[clip=on,width=10.5cm]{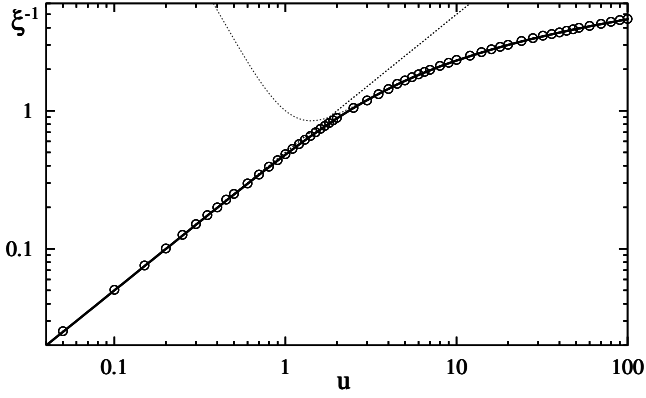}
\end{center}
\caption[]
{The inverse correlation length, $\xi^{-1}$, versus $u$
in $D=1$ crystal: the data obtained from high-accuracy computation
of ${\cal {R}}(\bm r)$ given by (\ref{Rhypergeo}) -- circles,
exact analytical result (\ref{ksi1}) -- thick continuous curve,
the asymptotic behavior of $\xi^{-1}$,
for small and large $u$, given in (\ref{ksi1asymp}) -- dotted curves.}
\label{fig1}
\end{figure}

Specifically, in $D=1$-case, the function ${\cal {R}}( r )$ at even points
($\sigma_r=r=2m$) can be expressed by
the Legendre function of the second kind, $Q_{\nu}(x)$: 
\bea
\label{Rhypergeo}
{\cal R}(2m)\!=\!\!
\frac{\kappa}{4\pi}\!\!\int\limits_0^{\pi}\!\!
\frac{\cos(2mk)\mbox{d}k}{\sqrt{\!1\!-\!\!\kappa^2\sin^2\!k}} 
\!=\!{\frac{(-1)^m}{2\pi}Q_{m-\frac{1}{2}}(1\!+\!\!\frac{u^2}{2})} ,
\eea
where $\kappa^2=(1+(u/2)^2)^{-1}$.
For sufficiently large $m$, the above expression can be cast in
the form (\ref{asymptot}), where
\bea
\label{ksi1}
\xi^{-1} \equiv
\ln ( \sqrt{1+ \ld {u}/{2} \rd ^2}+ {u}/{2} ), \quad \gamma={1}/{2} .
\eea
The small and large $u$ asymptotic behaviors of $\xi$ in (\ref{ksi1}) read:
\bea
\label{ksi1asymp}
\xi^{-1} \stackrel{u\rightarrow 0}{\approx} {u}/{2}+\ldots,  \ \
\mbox{and} \ \
\xi^{-1} \stackrel{u\rightarrow \infty}{\approx} \ln u + u^{-2} -\ldots .
\eea
The inverse correlation length given by the formulae
(\ref{ksi1}), (\ref{ksi1asymp}),
and by high-accuracy numerical calculations, plotted versus $u$,
are compared in Fig.~\ref{fig1}.
The asymptotic behavior (\ref{ksi1asymp}) has been also obtained
numerically in one-dimensional crystals with periodic potentials
of period greater than two  \cite{jkd}.

Naturally, in $D=2$-case the form of ${\cal {R}}( r )$  is more
complex \cite{jk}:
\bea
\label{R2hypergeom}
{\cal {R}}({\bm r})=
\frac{1}{\pi^2}\int_0^{\frac{\pi}{2}} \int_0^{\frac{\pi}{2}}{\mbox d}x{\mbox d}y
\frac{\cos(2nx)\cos(2my)}{\sqrt{\beta^2+\cos^2x\cos^2y}} \non
=\frac{(-1)^m}{2\pi} \frac{1}{\ld 2\beta \rd ^{2m+1}}
\frac{\Gamma^2(m+1/2)}{\Gamma(m+n+1)\Gamma(m-n+1)} \non
\times F\ld m+{1}/{2},m+{1}/{2};m+n+1;-{1}/{a} \rd \ \non
\times F\ld m+{1}/{2},m+{1}/{2};m-n+1;-{1}/{a} \rd ,
\eea
where $F(a,b;c;x)$ stands for the Gauss hypergeometric function, $\Gamma(a)$
for the Euler $\Gamma$-function, and the following notation was used
\bea
a\equiv2\beta(\sqrt{1+\beta^2}+\beta), \qquad \beta \equiv u/4, \non
r_1+r_2=2m=\sigma_{\bm r}, \ \ r_1-r_2=2n.
\eea
To analyse ${\cal {R}}({\bm r})$ given by (\ref{R2hypergeom}),
we first consider the diagonal direction ($r_1=r_2, n=0$).
Then, ${\cal {R}}({\bm r})$ can be expressed by the Legendre
function of first kind \cite{jk,as}, $P^m_{\nu}(x)$ :
\bea
\label{R20hypergeom}
{\cal {R}}({\bm r})=
\frac{(-1)^m}{4\pi^3\beta}\frac{1}{\Gamma^2(m+1/2)}
\lbr P^m_{-\frac{1}{2}}\ld 1+{2}/{a} \rd  \rbr^2 .
\eea
For sufficiently large $\sigma_{\bm r}=2m$, the above
${\cal {R}}({\bm r})$ assumes the form (\ref{asymptot}):
\bea
\label{R20asymp}
{\cal {R}}({\bm r})\stackrel{m \rightarrow \infty}{\approx}
\frac{(-1)^m}{2\pi\beta} \frac{\exp ( -2m/\xi )}{2m}
\eea
with
\bea
\label{ksi20}
\xi^{-1}=\ln(\sqrt{1+\beta^2}+\beta),   \qquad \gamma=1 .
\eea
Note that the inverse correlation length in (\ref{ksi20}) can be obtained
from that in $D=1$-case by a change of energy scale, $u \rightarrow u/2$.
Therefore, the small and large $u$ asymptotic behaviour of $\xi^{-1}$
is given by (\ref{ksi1asymp}) with the suitable change of the
energy scale. Consequently, in $D=1$-crystal and in the diagonal direction
of $D=2$-crystal, the inverse correlation length scales linearly with the
direct-gap width, $\xi^{-1} \sim \delta$, for sufficiently small $\delta$.

Second, we choose the direction along an axis, say the first axis
($r_2=0$, $\sigma_r=r_1=2m=2n$).
Then, the function  ${\cal {R}}({\bm r})$ in (\ref{R2hypergeom})
transforms into \cite{jk,as}:
\bea
\label{R2nhypergeom}
{\cal {R}}({\bm r})=
\frac{(-1)^m}{\pi}Q_{m-\frac{1}{2}}\ld 1+2a \rd
P_{m-\frac{1}{2}} \ld \frac{a-1}{a+1} \rd,
\eea
with the large $m$ asymptotics:
\bea
\label{R2nasymp}
{\cal {R}}({\bm r})\stackrel{m \rightarrow \infty}{\approx}
\frac{(-1)^m}{\pi\sqrt{2\beta}}\frac{\exp({-2m/\xi})}{2m} \non
 \times
 \cos\ld m\arccos \ld \frac{a-1}{a+1} \rd -\frac{\pi}{4}\rd ,
\eea
where the inverse correlation length depends on $u$ as follows:
\bea
\label{ksi2n}
\xi^{-1}\! \equiv\! \ln\!\!\lbr\! \sqrt{ \beta\!+\!\sqrt{1\!+\!\beta^2} }\ld  \!
\sqrt{2\beta}\!+\!\sqrt{\beta\!+\!\sqrt{1\!+\!\beta^2}}\rd \rbr ,
\eea
and $\gamma=1$.
Along an axis, the first terms of the small and large $u$ asymptotic
expansions of $\xi^{-1}$ read:
\bea
\label{ksi2nasymp}
\xi^{-1}\stackrel{u\rightarrow 0}{\approx}
\sqrt{\frac{u}{2}}\ld 1+ \frac{1}{12}\frac{u}{2}- \ldots \rd , \non
\ \  {\mbox {and}} \ \
\xi^{-1}\stackrel{u\rightarrow \infty}{\approx} \ln u +\!\frac{3}{u^2}-\!\ldots .
\eea

\begin{figure}
\begin{center}
\includegraphics[clip=on,width=10.5cm]{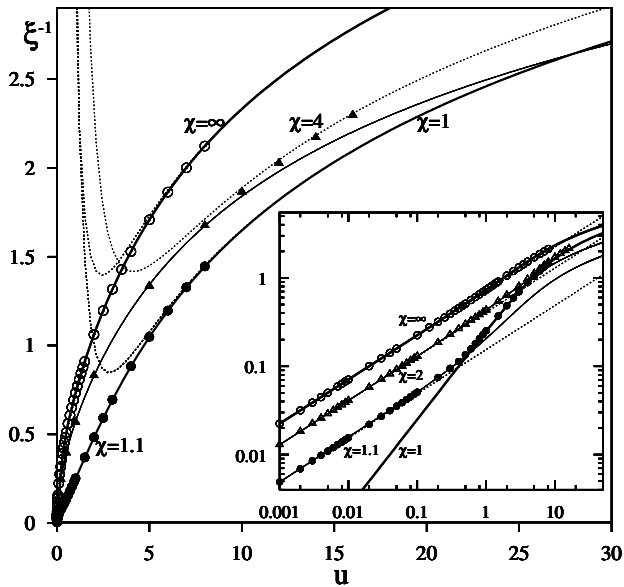}
\end{center}
\caption[]
{The inverse correlation length, $\xi^{-1}$, versus $u$,
for different directions $\chi$ in $D=2$ crystal:
the data obtained from high-accuracy computation
of ${\cal {R}}(\bm r)$ -- symbols,
exact analytical result along the diagonal (\ref{ksi20}) and
along an axis (\ref{ksi2n}) -- thick continuous curve,
exact analytical result along a direction given by slope $\chi$
(\ref{ksi2nm}) -- thin continuous curves,
exact analytical results for large $u$, given by
(\ref{ksi2nasymp}) and (\ref{ksi2nmasymplarge}) -- dots.
In the insert we show the log-log plot of $\xi^{-1}$ versus $u$,
and the dotted curve represents the first term of small $u$
asymptotics (\ref{ksi2nmasympsmall}).}
\label{fig2}
\end{figure}
Finally, we consider a general non-axial direction.
Let, for definitness, $0 < n < m$.
In this case we have not been able to obtain a large $m$ asymptotics of
the function  ${\cal {R}}({\bm r})$ in (\ref{R2hypergeom}) for arbitrary
values of $u$. However, we succeeded in deriving the asymptotic form
(\ref{asymptot}) for sufficiently small and sufficiently large values of $u$.
For sufficiently small values of $u$, we have found $\gamma=1$ and the inverse
correlation length
\bea
\label{ksi2nm}
\xi^{-1}\!\!=\!-
\!\frac{1}{2}\!\lb\! (1\!\!-\!\!\eta)\!\ln (\!\sqrt{1\!+\!\beta^2}\!+\!\!\beta)\! +
\!\!\ln\!\! \lbr \frac{2}{b\!+\!2}\!\ld \!\! 1\!-\!\!
\frac{2a}{b}\!\rd^{\eta} \rbr\!\! \rb\!\!, \
\eea
where the following notation has been used:
\bea
b\equiv(1+\eta )a+\sqrt{(1+\eta )^2a^2+4a\eta } \, , \non
\eta \equiv {n}/{m}=({\chi-1})/({\chi+1}) , \qquad  \chi\equiv {r_1}/{r_2}.
\eea
Around zero, the asymptotic expansion of the above $\xi^{-1}$ has the form
\bea
\label{ksi2nmasympsmall}
\xi^{-1}\stackrel{u\rightarrow 0}{\approx}
\sqrt{\frac{u}{2}\eta}
\ld 1+\frac{1}{24}\frac{1+\eta^2}{\eta}\frac{u}{2}-\ldots \rd.
\eea
Apparently, a straightforward way to obtain a large $\sigma_{\bm r}$
asymptotics of ${\cal {R}}(\bm r)$ is to expand $\Delta(\bm k)$
in powers of $u^{-2}$, for large $u$, carry out the integrals of products
of cosine functions and to approximate the Euler $\Gamma$-functions, that
arise, by the leading term of the Stirling's asymptotic expansion.
Such a procedure leads to the large-distance asymptotics of
${\cal {R}}(\bm r)$ (analogous to the one derived in \cite{taraskin1}),
which consists of the power-law factor with $\gamma=1$ and the exponential
factor with the inverse correlation length of the form
\bea
\label{ksi2nmasymplarge}
\xi^{-1}=\ln u+ (2+\chi+\chi^{-1})u^{-2} \non
-\sum_{\alpha=-1,1} (1+\chi^{\alpha})^{-1} \ln(1+\chi^{\alpha}).
\eea
In Fig.~\ref{fig2}, we have displayed the analytic results for
$\xi^{-1}$ versus $u$, obtained for $D=2$-crystal, and the same function
determined by means of high-accuracy numerical calculations.

To summarize, our work has  been inspired by the recent results of
Taraskin et al \cite{taraskin1}, concerning the spatial decay of DM
in a translation invariant model of insulator, where the relation
between the decay and the direct-gap width has not been established.
To the best of our knowledge, we provide in this letter the first
example of insulator being a crystal (with broken translational symmetry)
of dimension greater than one, where the relation between decay properties
of DM and the direct-gap width can be established exactly, in the whole
range of this parameter. Specifically, for arbitrary direct-gap width
the large-distance asymptotics of DM consists of the power law factor
with the power $\gamma = D/2$ and an exponential factor with the
correlation length depending on direction and the direct-gap width.

There is an excellent agreement between our
analytic results and high-accuracy numerical results presented
in the figures.
It is worth to emphasize that, by the general arguments given
in \cite{taraskin1}, these results should not depend on the
underlying lattice.

As particularly interesting, we consider the results concerning the
scaling of the correlation length $\xi$ with the vanishing
direct-gap width $\delta$, which is physically the most interesting regime,
in the $D=2$-crystal:
$\xi^{-1} \sim \delta$ in the diagonal direction while
$\xi^{-1} \sim \sqrt{\delta}$ in non-diagonal directions.
They resolve the controversy that has arisen in the literature
in this respect \cite{baer,beigi,taraskin1}:
which of these two scalings holds in insulators?
Apparently, the answer is more subtle than the question asked: both,
depending on lattice direction.

T.K. is grateful to the Institute of Theoretical Physics of the
University of Wroc\l aw for kind hospitality and financial support.


\begin{thebibliography}{}


\bibitem{payne}
M.C. Payne, M.P. Teter and D.C. Allan, T.A. Arias and J.D. Joannopoulos,
Rev. Mod. Phys. {\bf 64}, 1045 (1992)

\bibitem{goed}
S. Goedecker,
Rev. Mod. Phys. {\bf 71}, 1085 (1999)

\bibitem{kohn1}
W. Kohn,
Phys. Rev. Lett. {\bf 76}, 3168 (1996)

\bibitem{kohn2}
W. Kohn,
Phys. Rev. {\bf 115}, 809 (1959)

\bibitem{cloizeaux}
J. Des Cloizeaux,
Phys. Rev. {\bf 135}, A698 (1964)

\bibitem{baer}
R. Baer and M. Head-Gordon,
Phys. Rev. Lett. {\bf 79}, 3962 (1997)

\bibitem{beigi}
S. Ismail-Beigi and  T. A. Arias,
Phys. Rev. Lett. {\bf 82}, 2127 (1999)

\bibitem{he}
L. He and D. Vanderbilt,
Phys. Rev. Lett. {\bf 86}, 5341 (2001)

\bibitem{taraskin1}
S. N. Taraskin, D. A. Drabold, and S. R. Elliott,
Phys. Rev. Lett. {\bf 88}, 196405 (2002)

\bibitem{jkd}
J. J\c{e}drzejewski, T. Krokhmalskii and O. Derzhko,
arXiv:cond-mat/0211330

\bibitem{jk}
J. J\c{e}drzejewski and T. Krokhmalskii,
in preparation for publication

\bibitem{as}
M. Abramowitz and I. A. Stegun,
{\em Handbook of mathematical functions},
National Bureau of Standards, Applied Mathematics Series 55, 1964

\end{thebibliography}
\end{document}